\documentclass[aps,prb,twocolumn,showpacs,superscriptaddress,floatfix,nofootinbib]{revtex4-1} 
\usepackage{amsfonts}
\usepackage{graphicx}
\usepackage{amsmath}
\usepackage{times}
\usepackage{amssymb}
\usepackage{color}
\usepackage[colorlinks,bookmarks=false,citecolor=blue,linkcolor=red,urlcolor=blue]{hyperref}

\begin{document}
\title{$U(1)$-symmetric infinite projected entangled-pair state study of the spin-$1/2$ square $J_{1}-J_{2}$ Heisenberg model}
\author{R. Haghshenas}
\affiliation{Department of Physics and Astronomy, California State University, Northridge, California 91330, USA}
\author{D. N. Sheng}
\affiliation{Department of Physics and Astronomy, California State University, Northridge, California 91330, USA}
\date{\today}
\begin{abstract}
We develop an improved variant of $U(1)$-symmetric infinite projected entangled-pair state (iPEPS) ansatz to investigate the ground state phase diagram of the spin-$1/2$ square $J_{1}-J_{2}$ Heisenberg model. In order to improve the accuracy of the ansatz, we discuss a simple strategy to select automatically relevant symmetric sectors and also introduce an optimization method to treat second-neighbor interactions more efficiently.  We show that variational ground-state energies of the model obtained by the $U(1)$-symmetric iPEPS ansatz (for a fixed bond dimension $D$) set a better upper bound, improving previous tensor-network-based results. By studying the finite-$D$ scaling of the magnetically order parameter, we find a N\'{e}el phase for $J_2/J_1<0.53$. For $0.53<J_2/J_1<0.61$, a non-magnetic columnar valence bond solid (VBS) state is established as observed by the pattern of local bond energy. The divergent behavior of correlation length $\xi \sim D^{1.2}$ and vanishing order parameters are consistent with a deconfined N\'{e}el-to-VBS transition at $J^{c_1}_{2}/J_1=0.530(5)$, where estimated critical anomalous exponents are $\eta_{s} \sim 0.6$ and $\eta_{d} \sim 1.9$ for spin and dimer correlations respectively. We show that the associated VBS order parameter monotonically increases with $J_2/J_1$ and finally a first-order quantum phase transition takes place at $J^{c_2}_{2}/J_1=0.610(2)$ to the conventional Stripe phase. We compare our results with earlier DMRG and PEPS studies and suggest future directions for resolving remaining issues.
\end{abstract}

\maketitle
\section{Introduction}
\label{Sec:introduction}
Understanding of quantum many-body systems is of fundamental importance. These systems, even in the simplest form, reveal fascinating quantum collective behavior distinctly different from noninteracting particles. For instance, frustrated quantum spin systems, defined by a simple spin model, are considered one of the most important playgrounds to observe exotic phenomena. Quantum spin liquid\cite{Balents:2010, savary:2016} with a topologically order,\cite{Wen:1990, Read:1991} valence bond solid (VBS) order \cite{Affleck:1987, read:1989, read:1990} and deconfined quantum criticality \cite{Senthil:2004} are some of well-known examples manifested in such systems. Specifically, searching for the quantum spin-liquid states has received much attention due to their distinct characteristics, such as long-range entanglement \cite{Chen:2010} and nontrivial anyon statistics.\cite{Wen:1990, Wen:2007} A comprehensive characterization of them might lead to new understanding in physics of frustrated quantum magnetism and providing `a new theoretical framework' \cite{Chen:2011:Jan} for characterizing exotic phases of matter.

$J_{2}\sim J_{1}$ The frustrated spin-$1/2$ $J_{1}-J_{2}$ square Heisenberg model (SHM) is one of the candidate models featuring aforementioned exotic phases. The $J_{1}-J_{2}$ SHM has stimulated extensive theoretical studies over the last two decades, due to its simplicity and its experimental realization in several materials, \cite{Melzi:2000, Nath:2008, Koga:2016} such as vanadium Layered oxides ${\text{Li}}_{2}\text{VO}(\text{Si,Ge}){\text{O}}_{4}$ and polycrystalline samples $\text{BaCdVO}{({\text{PO}}_{4})}_{2}$. In particular, these studies have established that the second-neighbor $J_{2}$ coupling controlling frustration induces non-magnetic phases around the highly frustrated point $J_2/J_1=0.5$.\cite{ IOFFE:1988,Zhitomirsky:1996, Capriotti:2000, Takano:2003, Mambrini:2006, Capriotti:2001, Zhang:2003,Sirker:2006,Darradi:2008, Beach:2009,Richter:2010,Wang:2013, Sachdev:1990, Singh:1999, Capriotti:2001, Chubukov:1991, Murg:2009, Sadrzadeh:2016} Despite that, depending on numerical approaches, several scenarios have been proposed around this point: the earlier studies based on small-size exact diagonalization, spin-wave theory, series expansion and coupled cluster methods find different candidate states, such as columnar,\cite{Sachdev:1990, Singh:1999,Capriotti:2001, Chubukov:1991} plaquette VBS states\cite{Zhitomirsky:1996, Capriotti:2000, Takano:2003,Mambrini:2006} and resonating valence bond \cite{Capriotti:2001, Zhang:2003} spin liquid states. 


The recent  $SU(2)$-symmetric density matrix renormalization group (DMRG) study has demonstrated an intermediate plaquette VBS phase between a N\'{e}el and Stripe magnetically ordered phases,\cite{Gong:2014} which does not support the previous DMRG results of gapped $\mathcal{Z}_{2}$ spin liquid as the intermediate phase.\cite{Hong-Chen:2012} However, in a small window of $0.44<J_2/J_1 \leq 0.5$, the $SU(2)$ DMRG results cannot distinguish between two possible scenarios, between a true deconfined quantum-critical point or a gapless spin-liquid phase. A very recent DMRG study \cite{Wang:2017} further supports the possibility of a gapless spin liquid between the N\'{e}el and the VBS phases by following the energy level crossings between different low energy excited states. On the other hand, variational Monte Carlo (VMC) results \cite{Hu:2013} predict a gapless $\mathcal{Z}_{2}$ spin liquid in the whole region $0.45 \leq J_{2}/J_1 \leq 0.6$, while a very recent VMC study \cite{Morita:2015} challenged this result by predicting a columnar VBS order for $0.5\leq J_{2}/J_1 \leq 0.6$. The critical exponents reported in this study show small deviation from those of the $\mathcal{J}$-$\mathcal{Q}$ models. However, understanding the true nature of quantum critical points and the corresponding universality classes turn out to be even more challenging using unbiased methods.\cite{Hong-Chen:2012, Sandvik:2012, Gong:2014}

Recently, tensor-network-based methods have also been applied to study the $J_{1}-J_{2}$ SHM. An early plaquette renormalized tensor-network study \cite{Wang:2011} has predicted a possible plaquette VBS order for the intermediate phase. They estimated the second-order phase transition between N\'{e}el and plaquette VBS phase to occur around $J^{c_{1}}_{2}/J_1\approx 0.40$. On the other hand, finite-size projected entangled pair states (PEPS) ansatz with the cluster-update scheme \cite{Lubasch:2014} finds a direct N\'{e}el-to-VBS transition occurring at $J^{c_{1}}_{2}/J_1\approx0.57$.\cite{Wang:2016} The finite-size PEPS results did not identify the true nature of VBS order, specifically between plaquette and columnar. They also find corresponding critical exponents are consistent with the $\mathcal{J}$-$\mathcal{Q}$ models. A very recent $SU(2)$-symmetric infinite PEPS (iPEPS) ansatz suggests a quantum critical point at $J^{c_{1}}_{2}/J_1 \simeq0.5$, where in contrast to the finite-size PEPS results,\cite{Wang:2016} the extracted critical exponents seem to deviate from those of the $\mathcal{J}$-$\mathcal{Q}$ models.\cite{Poilblanc:2017}

In this paper, we aim to develop a fully $U(1)$-symmetric iPEPS ansatz with an `improved' update scheme to reexamine the phase diagram of the $J_{1}-J_{2}$ SHM. So far, the iPEPS update algorithms \cite{Corboz:2010:April, Corboz:2010} have been able to treat the first-neighbor interactions with high efficiency. They have been shown in practice to be quite accurate and stable providing reliable results. However, in the case of longer-range interactions (e.g. second-neighbor interactions) a similarly efficient scheme is still highly desired. To this end, we present a new update method based on the so-called positive approximant and reduced-tensor application \cite{Lubasch:2014, Phien:2015} to treat second-nearest neighbor interactions more accurately and efficiently. We find that the new update scheme significantly improves efficiency and provides more accurate results in comparison with previous schemes.\cite{Corboz:2013, Corboz:2010} In addition, we also investigate the implementation of $U(1)$ symmetry into the iPEPS ansatz by introducing a general scheme to pick up relevant symmetry sectors. We show that it solves the loss of accuracy observed when applying continuous symmetry groups \cite{Bauer:2011} and provides the same accuracy as non-symmetric iPEPS.  


By using the $U(1)$-symmetric iPEPS ansatz, we clarify the quantum phase diagram and the nature of phase transitions for the $J_1$-$J_2$ SHM with substantially improved variational wave function (of the ground state), and bridge the gap between the previous tensor-network and DMRG studies. We show that the non-magnetic phase appears in the range of $0.53 < J_{2}/J_1\leq 0.61$. The critical point $J^{c_1}_{2}/J_1\simeq0.53$ is of the deconfined type confirmed by continuously vanishing the N\'{e}el order parameter and the divergence of the correlation length $\xi \sim D^{1.2}$. By extrapolating dimer-dimer and spin-spin correlation functions in the $D \rightarrow \infty$ limit, we estimate the critical anomalous exponents $\eta_{s} \sim 0.6$ and $\eta_{d} \sim 1.9$. The pattern of the local nearest neighboring bond energies shows that a columnar VBS phase is established up to $J^{c_2}_{2}/J_1\simeq0.61$. However, the observed (variational) energies from different approaches \cite{Gong:2014} indicate both columnar and plaquette VBS phases are competitive candidates for the intermediate phase. With further increasing $J_2/J_1$, a first-order phase transition takes place from VBS phase to the conventional Stripe phase. 


The paper is organized as follows. We first introduce the model and briefly summarize different types of the phases and the resulting phase diagram obtained by our iPEPS studies in Sec.~\ref{Sec:Model}. In Sec.~\ref{Sec:Methods}, we briefly introduce the $U(1)$-symmetric iPEPS ansatz and discuss a general scheme to select automatically relevant symmetric sectors (Sec.~\ref{Sec:sectors}). We then present a new iterative scheme in detail and compare it with previous schemes (Sec.~\ref{Sec:Optimization}). Sec.~\ref{Sec:Results} provides the main simulation results. The variational ground-state energy and N\'{e}el order parameter are presented in Secs.~\ref{Sec:Neelphase}. We show that the intermediate phase is a columnar VBS represented in Sec.~\ref{Sec:ColumnarVBSphase}. The critical properties of the deconfined quantum-critical point are discussed by studying correlation function and correlation length in Sec.~\ref{Sec:Deconfined}---further plots of the correlation functions are presented in Appendix.~\ref{App:Extrapolated}. Using different initial tensors representing different symmetry breaking states, we determine the boundary of columnar VBS and the conventional Stripe phase in Sec.~\ref{Sec:FirstOrder}. Finally, we summarize our work with some discussions in Sec.~\ref{Sec:CONCLUSION}.

\section{Model}
\label{Sec:Model}
The $J_{1}-J_{2}$ SHM is defined by the Hamiltonian
\begin{equation*}
H=J_{1} \sum_{\langle i,j\rangle}\textbf{S}_{i}\cdot\textbf{S}_{j}
+J_{2}\sum_{\langle\langle i,j\rangle\rangle}\textbf{S}_{i}\cdot\textbf{S}_{j},
\end{equation*}
where ${\textbf{S}_i}\equiv(\textbf{S}_i^x, \textbf{S}_i^y, \textbf{S}_i^z)$ are spin-$1/2$ operators. The couplings $J_1$ and $J_2$ stand for the first- and second-neighbor antiferromagnetic (AFM) interactions. We set $J_{1}=1$ throughout the paper and consider the frustrated interaction $J_{2}>0$.  

\begin{figure}
\begin{center}
\includegraphics[width=1.0 \linewidth]{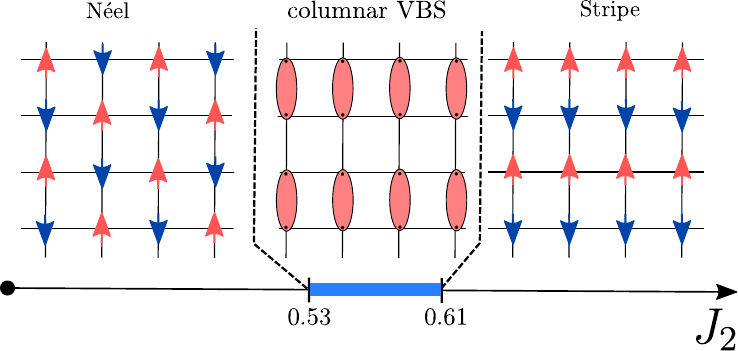} 
\caption{ (Color online) Phase diagram of  the $J_{1}-J_{2}$ SHM as a function of coupling $J_{2}$. The arrows show pattern of magnetic order appeared in AFM N\'{e}el and Stripe phases. The eclipses in intermediate phase (columnar VBS) stand for entangled spins (singlet states).}
\label{fig:phasediagrame}
\end{center}
\end{figure}

In the extreme cases $J_{2} \approx 0$ or $J_{2} \gg 1$, the ground states are respectively defined by two magnetically ordered phases, i.e., AFM N\'{e}el and Stripe. The patterns of magnetic orders for these phases have been shown in Fig.~\ref{fig:phasediagrame}. All the earlier studies suggest that these two phases are separated by an (or several) intermediate phase(s). Our goal is to locate and characterize the intermediate phase.

The obtained phase diagram has been illustrated in Fig.~\ref{fig:phasediagrame}. We find that the intermediate phase is a paramagnetic phase that breaks lattice symmetry, i.e., a columnar VBS. As seen in Fig.~\ref{fig:phasediagrame}, columnar VBS order (in which vertical spins are strongly entangled) only breaks lattice symmetry in the $y$-direction. The columnar VBS phase is separated from the N\'{e}el one by a continuous phase transition occurred at $J^{c_1}_{2}=0.530(5)$. In addition, the quantum phase transition between VBS and AFM Stripe phases takes place at $J^{c_2}_{2}=0.610(3)$, which is of the first-order type. 

\section{METHOD}\label{Sec:Methods}
\subsection{$U(1)$-symmetric iPEPS ansatz}
\label{Sec:ipeps}
\begin{figure}
\begin{center}
\includegraphics[width=0.90 \linewidth]{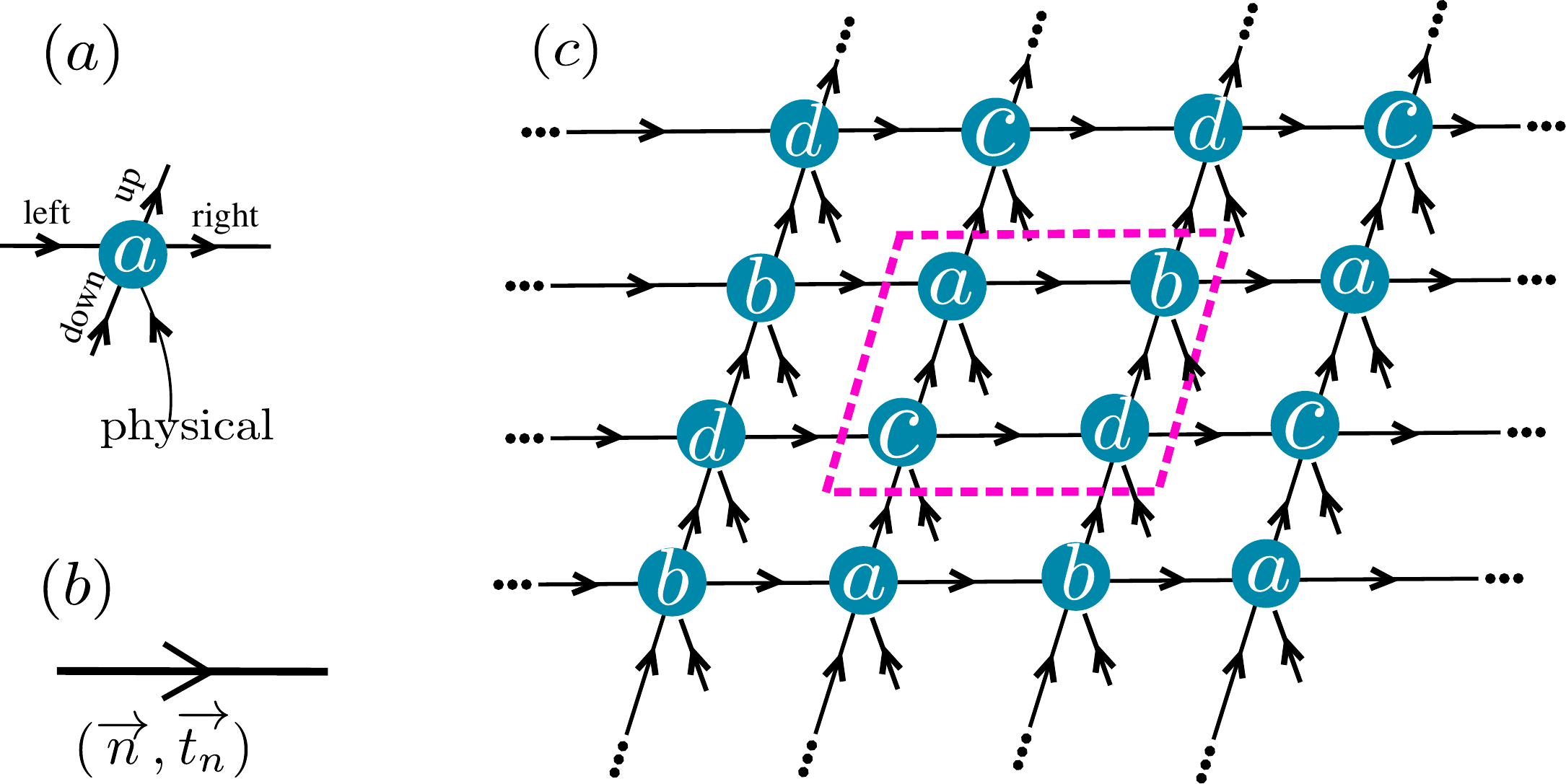} 
 \caption{(Color online) Tensor-network representation of the iPEPS ansatz. $(a)$ $U(1)$-invariant five-rank tensor (particle numbers associated to incoming and outgoing arrows are equal). Virtual bonds are labeled by $\{\text{left},\text{down}, \text{right}, \text{up}\}$. $(b)$ A virtual bond is labeled by vectors $(\protect\overrightarrow{n}, \protect\overrightarrow{t_{n}})=(\cdots, (n^{(i)}, t^{(i)}_{n}), \cdots)$, where $i$th components $n^{(i)}$ and $t^{(i)}_{n}$ represent a particle number and its associated dimension.  $(c)$ $U(1)$-symmetric $2 \times 2$  unit cell iPEPS. }
  \label{fig:ipeps}
\end{center}
\end{figure}

An iPEPS is constructed by building-block tensors that are sitting on sites of the physical lattice.\cite{Verstraete:2008} The tensors are connected to each other by the so-called virtual bonds (graphically drawn by arrows) constructing a specific geometrical pattern (usually similar to physical lattice). For instance, as depicted in Fig.\ref{fig:ipeps}-(a-c), we have constructed a $2 \times 2$ unit cell iPEPS on the infinite two-dimensional square lattice by repeating periodically five-rank tensors $\{a, b, c, d\}$. The geometrical structure produced by the connections of tensors has an important feature: the iPEPS could reproduce entanglement area law.\cite{Verstraete:2008, Orus:2014} The amount of this entanglement is controlled by the so-called bond dimension (number of elements) of virtual bonds, denoted by $D$. By increasing $D$, the iPEPS is able to represent highly entangled stats. 

We aim to use the iPEPS as a variational ansatz to obtain the approximate ground state of the model. The accuracy of this variational method is controlled by the bond dimension $D$ (variational parameters are of order $\mathcal{O}(D^{4})$). To capture the physics of highly entangled states, one needs to consider larger $D$ and does the finite-$D$ analysis (extrapolating $D \rightarrow \infty$ ). By exploiting $U(1)$ symmetry, one can study the iPEPS with a larger $D$. In the presence of this symmetry, each tensor takes a block diagonalized form (each block is corresponding to a specific symmetric sector) which correspondingly reduces computational costs.\cite{Singh:2010} The symmetric sectors in the case of $U(1)$ symmetry are labeled by the conserved particle numbers $\protect\overrightarrow{n}$. To implement this symmetry into the iPEPS, we label each arrow by some particle numbers $\protect(\overrightarrow{n},\overrightarrow{t_n})=(\cdots, (n^{(i)}, t^{(i)}_{n}), \cdots)$, as depicted in Fig.\ref{fig:ipeps}-(b), so that their sign ($\pm$) being specified by outgoing and incoming arrows.\cite{Singh:2011} For example, a virtual bond with the associated label $\protect(\overrightarrow{n},\overrightarrow{t_n})=((-1,2), (1,3))$ would have the bond dimension $D=5$ with particle numbers $-1, 1$ and associated dimensions $2, 3$ respectively.  Each tensor is $U(1)$ invariant as the sum of incoming particle numbers equals to the sum of outgoing ones. In this case, when each individual tensor is $U(1)$ invariant, the iPEPS automatically respects $U(1)$ symmetry.

\subsection{Selection of relevant symmetric sectors}
\label{Sec:sectors}
For infinite symmetric groups, like $U(1)$, it is not possible to uniquely specify symmetric sectors $\protect\overrightarrow{n}$. In addition, number of states in each sector $\protect\overrightarrow{t_{n}}$ should be manually chosen. Furthermore, the virtual bonds could possess non-homogeneous structures: each virtual bond takes different symmetric sectors from another one. These possibilities in selecting the symmetric sectors impede the $U(1)$-symmetric iPEPS ansatz from providing accurate results as reported in Ref.~\onlinecite{Bauer:2011}---note, this loss of accuracy is not observed in the case of finite symmetric groups (even for homogeneous-bond structure) due to the finite number of the symmetric sectors. 
\begin{figure}
\begin{center}
\includegraphics[width=1.0 \linewidth]{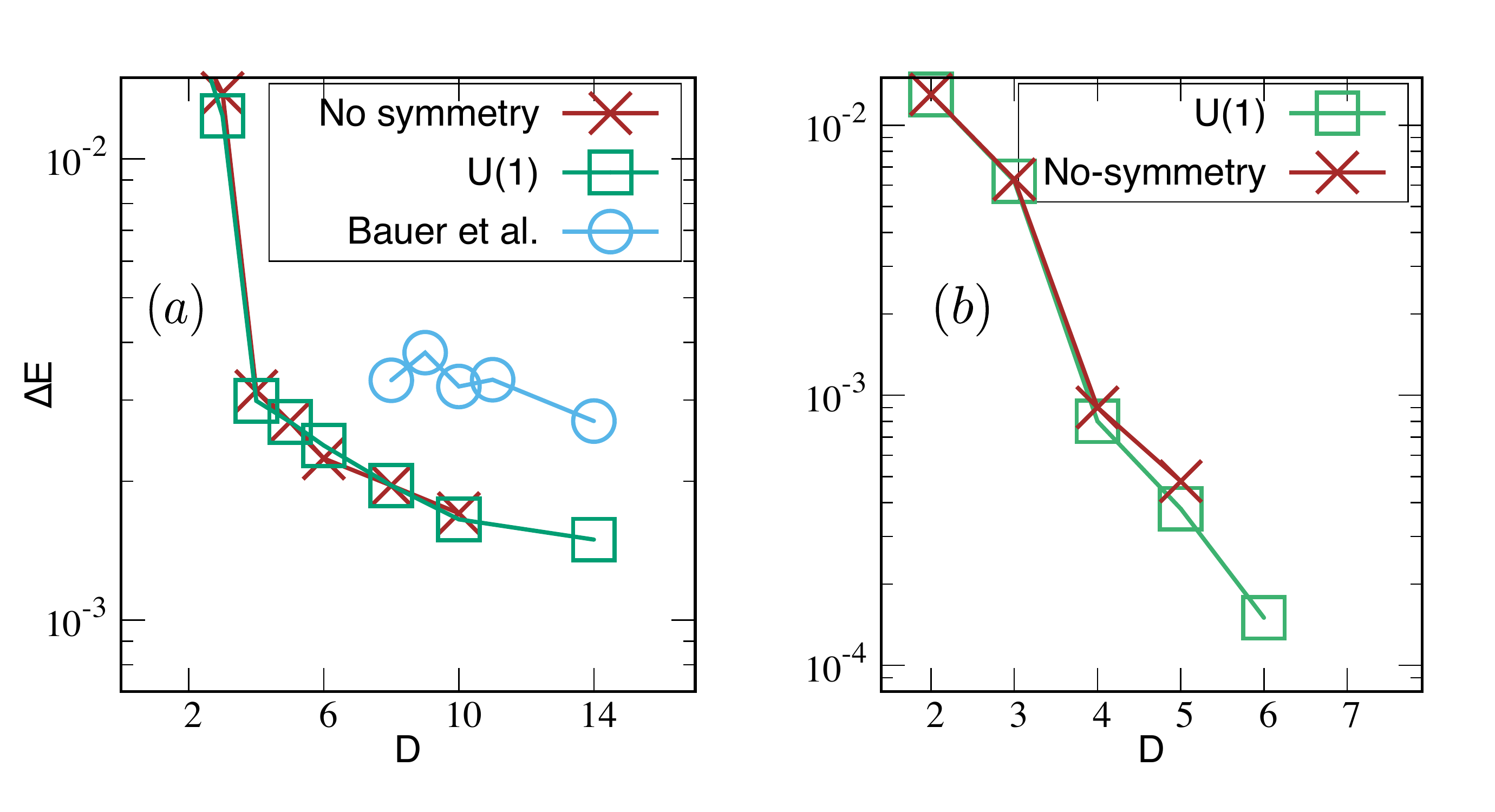}  
  \caption{(Color online) Relative error $\Delta E$ in the ground-state energy of Heisenberg model on square lattice ($J_{2}=0$) for $(a)$ simple- and $(b)$ full-update schemes. The data shown by blue circles are obtained by homogeneous $U(1)$-symmetric iPEPS (all virtual bond have the same symmetry sector) reported in Ref.~\onlinecite{Bauer:2011}. The green square symbols show our results (a non-homogeneous structure, see for example Tab.~\ref{tab:symmetric}) obtained by the scheme explained in the main text.} 
\label{fig:SU}
\end{center}
\end{figure}

We introduce a simple strategy to select automatically relevant symmetric sectors $\protect(\overrightarrow{n},\overrightarrow{t_n})$ by using simple-update simulation.\cite{Jiang:2008, Gu:2008, Corboz:2010:April} In this scheme, we assume a non-homogeneous structure for the virtual bonds (each of them could take different symmetric sectors). We then perform simple-update simulation: $(i)$ randomly initialize iPEPS (picking up a random set of symmetric sectors), $(ii)$ apply the local imaginary time-evolution operator \cite{Corboz:2010} to the virtual bonds and $(iii)$ use high-order singular value decomposition \cite{Xie:2012} to keep the largest singular values, which also determines the symmetric sectors. In addition, to obtain the expectation values, we similarly assume a non-homogeneous structure for the so-called environment tensors \cite{Phien:2015, Orus:2012, Orus:2009} and pick up the symmetric sectors by using the singular values decomposition appeared in the corner transfer matrix (CTM) approach.\cite{Nishino:1997} Furthermore, to do the full-update simulation,\cite{Phien:2015} we first fix the symmetric sectors for all virtual bonds (obtained by the aforementioned scheme), then randomly initialize tensors, and finally apply the optimization schemes (e.g., as explained in Sec.~\ref{Sec:Optimization}).

\begin{table}
\begin{tabular}{|c|c|c|c|c|}
\hline $tensor$ &$\text{left}$ &$\text{down}$ &$\text{right}$& $\text{up}$ \\ \hline
$a$ &$(-2, 0, 2)$&$(-3, -1, 1)$ &$(-1,1, 3)$&$(-3, -1, 1)$ \\
$b$ &$(-1,1, 3)$&$(-1,1, 3)$&$(-2, 0, 2)$&$(-1, 1, 3)$ \\
$c$ &$(-1,1, 3)$ & $(-3, -1, 1)$&$(-2, 0, 2)$&$(-3, -1, 1)$ \\
$d$ &$(-2, 0, 2)$ & $(-1,1, 3)$&$(-1,1, 3)$&$(-1, 1, 3)$ \\ \hline
\end{tabular}
\caption{The resulting particle numbers $\protect\overrightarrow{n}$ for the virtual bonds of tensors $\{a, b, c, d\}$ for the Heisenberg model $J_{2}=0$. The associated degeneracy is always $\protect\overrightarrow{t_n}=(2, 3, 2)$ for all virtual bonds. The labels $\{\text{left},\text{down}, \text{right}, \text{up}\}$ show four virtual bonds of each tensor. The particle numbers have obtained by the scheme explained in the main text. }
\label{tab:symmetric}
\end{table}

We apply this scheme to the Heisenberg model on the square lattice ($J_{2}=0$) to compare its accuracy with that of previous ones presented in Ref.~\onlinecite{Bauer:2011}. We use the relative error $\Delta E$ in the ground-state energy to provide benchmarks: $\Delta E =\frac{E_{D}-E_{MC}}{E_{MC}}$, where $E_D$ and  $E_{MC}$ are respectively the iPEPS energy with finite bond-dimension $D$ and the precise Monte-Carlo energy from Ref.~\onlinecite{Sandvik:1997}. As shown in Fig.~\ref{fig:SU}-(a, b), a proper choice of symmetric sectors makes the $U(1)$ iPEPS highly accurate. We observe that a $U(1)$-symmetric iPEPS ansatz produces the same accuracy as the non-symmetric ones in both full- and simple-update methods for the same bond dimension $D$. The resulting symmetry sectors for virtual bonds (with $D=7$) are shown in Tab.~\ref{tab:symmetric}, where we have started from a homogeneous structure. The particle numbers $\protect\overrightarrow{n}$ dynamically vary during the simulation and (finally) take a non-homogeneous structure---as $\protect\overrightarrow{n}$ is different for each virtual bond. Specifically, our results imply that the non-homogeneous structures are crucial in the case of $U(1)$ symmetry. 


\subsection{Optimization method}
\label{Sec:Optimization}
In order to do a full-update simulation for the models including second-neighbor interactions, we introduce a new iterative scheme to optimize the tensors. We use both applications of positive approximant \cite{Phien:2015, Lubasch:2014} and reduced tensors \cite{Corboz:2010:April} in an iterative way to improve accuracy and convergence rate of the optimization algorithm. We discuss the general ideas here, but for computational implementation and more details see Refs.~\onlinecite{Kao:2015,Verstraete:2008,Orus:2014}.

To perform a full-update simulation, we need to study imaginary-time evolution of an initial (random) iPEPS $|\psi(a, b, c, d)\rangle$. We use first-order Suzuki-Trotter decomposition \cite{Suzuki:1990} to split the imaginary time-evolution operator into a sequence of local terms. Such local operators are acting on specific bonds, increasing the corresponding bond dimension $D \rightarrow D'$. In order to have a tractable algorithm, we need to reduce the bond dimension (approximating the resulting iPEPS). We explain this procedure by considering local imaginary-time operators acting on, e.g., tensors $\{a, b, c\}$
\begin{equation*}
U=e^{-\delta  (J_{1}\textbf{S}_{c}\cdot\textbf{S}_{a}+J_{2}\textbf{S}_{c}\cdot\textbf{S}_{b}+J_{1}\textbf{S}_{a}\cdot\textbf{S}_{b})  },
\end{equation*}      
where $\delta$ stand for small time steps and $\textbf{S}_{a}$ is acting on tensor $a$ (analogous for other operators). After applying $U$, the resulting wave function should be approximated by a new iPEPS with the bond dimension $D$
\begin{equation*}
| \psi'(a',b',c',d') \rangle \approx U|\psi(a, b, c, d)\rangle,
\end{equation*}
where tensors $\{a', b', c', d'\}$ are determined by minimizing the square distance. We consider tensors $\{a', b', c', d'\}$ as variational parameters and accordingly find them by minimizing the square distance
\begin{equation*}
\min_{\{a',b',c',d'\}} \, f(| \psi'((a',b',c',d')) \rangle, U|\psi(a, b, c, d)\rangle),
\end{equation*}
where 
\begin{eqnarray*}
f=\langle \psi|U^{\dagger}U|\psi\rangle+\langle \psi'|\psi'\rangle-\langle \psi'|U|\psi\rangle-\langle \psi|U^{\dagger}|\psi'\rangle. 
\end{eqnarray*}
To minimize the cost function $f$, we use positive approximant and reduced-tensor schemes:
     
\begin{figure}
\begin{center}
\includegraphics[width=0.90 \linewidth]{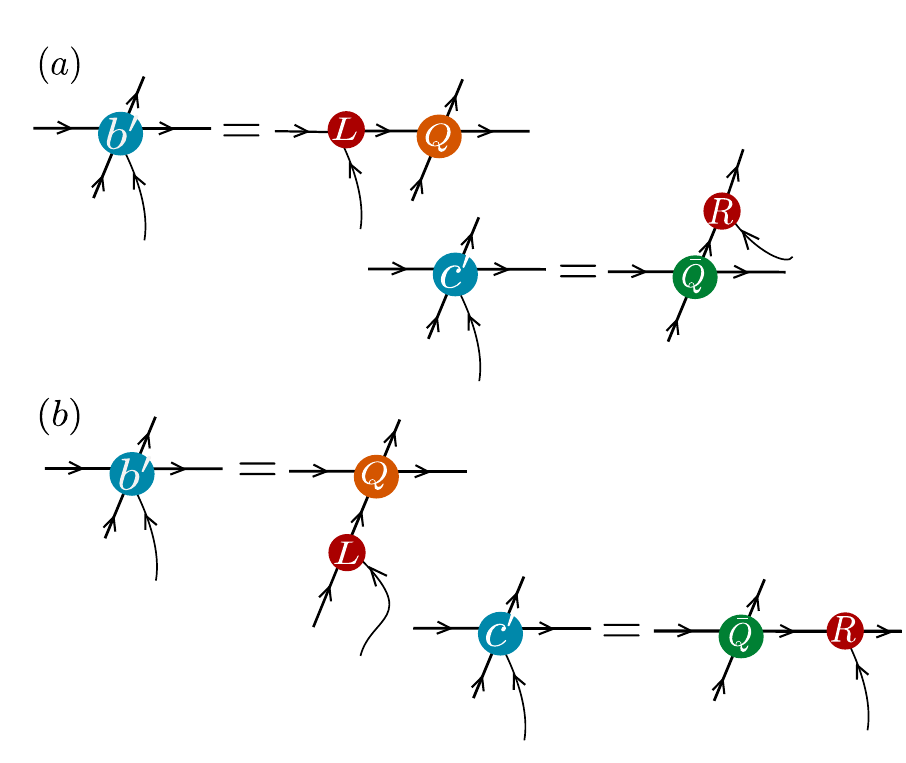} 
\caption{(Color online) Tensor-network diagram of reduced-tensor application. Tensors $\{b', c'\}$ are decomposed to low-rank tensors $\{l, r, Q, \bar{Q}\}$ by the means of LQ and QR decomposition. For example in $(a)$, tensors $L$ and $Q$ are obtained by fusing $\{\text{left}, \text{up}, \text{down}\}$ indices and $\{\text{right}, \text{physical}\}$ indices of tensor $b'$ (to make a matrix) and performing LQ decomposition of that matrix.}
\label{fig:QR}
\end{center}  
\end{figure}

\begin{figure}[t]
\includegraphics[width=0.80 \linewidth]{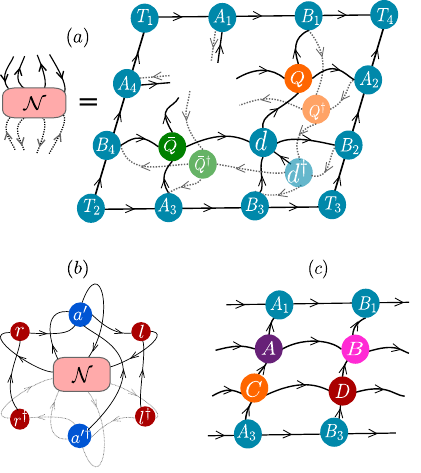} 
  \caption{(Color online) . $(a)$ Tensor-network representation of norm tensor $\mathcal{N}$. Tensors $\{A_{1},\cdots,A_{4}, B_{1},\cdots, B_{4}, T_{1},\cdots, T_{4}\}$ are the environment tensors (with bond dimenstion $\chi$) obtained by CTM renormalization group approch.\cite{Corboz:2014, Huang:2012} $(b)$ tensor-network representation of the term $r^{\dagger} a'^{\dagger} l^{\dagger} \mathcal{N}  l a' r$ appeared in cost function $f$. Note that by taking conjugate transpose the direction of arrows changes. $(c)$ Tensor-network representation of transfer matrix. For instance, tensor $A$ is obtained by contracting physical index of tensors $a^{\dagger}$ and $a$, and then fusing (combining) corresponding virtual bonds---so, the bond dimension of each virtual bond is $D^{2}$.} 
\label{fig:OptEXPCT}  
\end{figure}
     
\begin{itemize}
    \item[(a)] \underline{\it reduced-tensor application:} We use QR and LQ decomposition to split tensors $\{b', c'\}$ to sub-tensors $\{l, r, Q, \bar{Q}\}$ as depicted in Fig.~\ref{fig:QR}-(a). We aim to minimize the cost function with respect to tensors $\{l, a', r\}$, thus we rewrite the cost function as following
\begin{equation*}
\min_{\{a', r, l\}} \, f= const+r^{\dagger} a'^{\dagger} l^{\dagger} \mathcal{N}  l a' r - r^{\dagger} a'^{\dagger} l^{\dagger} \bar{\mathcal{N}}-\bar{\mathcal{N}^{\dagger}}ra'l,
\label{EQ:gaugefixing}
\end{equation*}
where $\mathcal{N}$ is called norm tensor. Tensor-network representations of the norm tensor and the term $r^{\dagger} a'^{\dagger} l^{\dagger} \mathcal{N}  l a' r$ are shown in Fig.~\ref{fig:OptEXPCT}-(a, b). Note that the first term does not play any role in the optimization procedure.

    \item[(b)] \underline{\it positive approximant:} In principle, the norm tensor $\mathcal{N}$ should be positive and Hermitian. But mainly due to the CTM approximation, it has some small negative parts. We explicitly eliminate that part by enforcing $\mathcal{N}$ to be positive. We also replace $\mathcal{N}$ by its Hermitian positive counterpart ($\mathcal{N_{+}}$) in the cost function: $\mathcal{N_{+}}=\sqrt{\mathcal{\overline{N}}^{2}}$, where $\mathcal{\overline{N}}=(\mathcal{N}+\mathcal{N}^{\dagger})/2$. 
    
    \item[(c)] \underline{\it alternating-least-squares (ALS) sweep:} We then iteratively optimize the cost function by finding the optimum tensors $\{l, a', r\}$: e.g., we minimize the cost function with respect to $a'$ by solving equation $\partial_{a'^{\dagger}} f=0$ by holding fixed tensors $l, r$. Then we repeat this procedure for another tensor with holding rest fixed until cost function converges.  

    \item[(d)] \underline{\it recovering:} After finding optimum tensors $\{l, a', r\}$, we absorb tensor $\{l, r\}$ to $\{Q, \bar{Q}\}$ to recover the final optimum tensors $\{b', a', c'\}$.
\end{itemize}

The positive approximant in our scheme is crucial in making the algorithm highly stable and accelerating its convergence. The computational cost of the norm tensor and ALS sweep are respectively $\mathcal{O}(D^{6}\chi^{3})$ and $\mathcal{O}(D^{12})$, where $\chi$ is the bond dimension of the environment tensors. Since we only need to calculate the norm tensor once, the dominant computational cost belongs to ALS sweep, i.e., $\mathcal{O}(D^{12})$. We should also notice in the case that $\chi>D^{2}$, that is suppressed by $\mathcal{O}(D^{6}\chi^{3})$ (as occurs in our calculations).

Although, in practice, the steps-($a$-$d$) provide proper accuracy and approximates the iPEPS wave function well, but there is still room to improve it. Specifically, we did not take into account tensor $d'$ in the optimization procedure (left untouched). In addition, sub-tensor application might reduce the accuracy. The main idea is to use sub-tensor application in a different way (see Fig.~\ref{fig:QR}-(b)) to design an efficient strategy to include tensor $d'$ in the optimization procedure. The steps are as follows: $(i)$ we decompose tensors $\{b', c'\}$ to sub-tensors $\{l, r, Q, \bar{Q}\}$ as shown in Fig.~\ref{fig:QR}-(b) and rewrite the cost function $f= const+r^{\dagger} d'^{\dagger} l^{\dagger} \mathcal{N}  l d' r - r^{\dagger} d'^{\dagger} l^{\dagger} \bar{\mathcal{N}}-\bar{\mathcal{N}^{\dagger}}rd'l$, $(ii)$ use positive approximation $\mathcal{N} \rightarrow \mathcal{N_{+}}$, $(iii)$ optimize the cost function by finding the optimum tensors $\{l, d', r\}$ and $(iv)$ after finding optimum tensors $\{l, d', r\}$, we absorb tensor $\{l, r\}$ to $\{Q, \bar{Q}\}$ to recover the final optimum tensors $\{b', d', c'\}$. The computational costs for the steps-($i$-$iv$) are similarly $\mathcal{O}(D^{6}\chi^{3})$ and $\mathcal{O}(D^{12})$.

%

\begin{figure}[b]
  \begin{center}
\includegraphics[width=1.04 \linewidth]{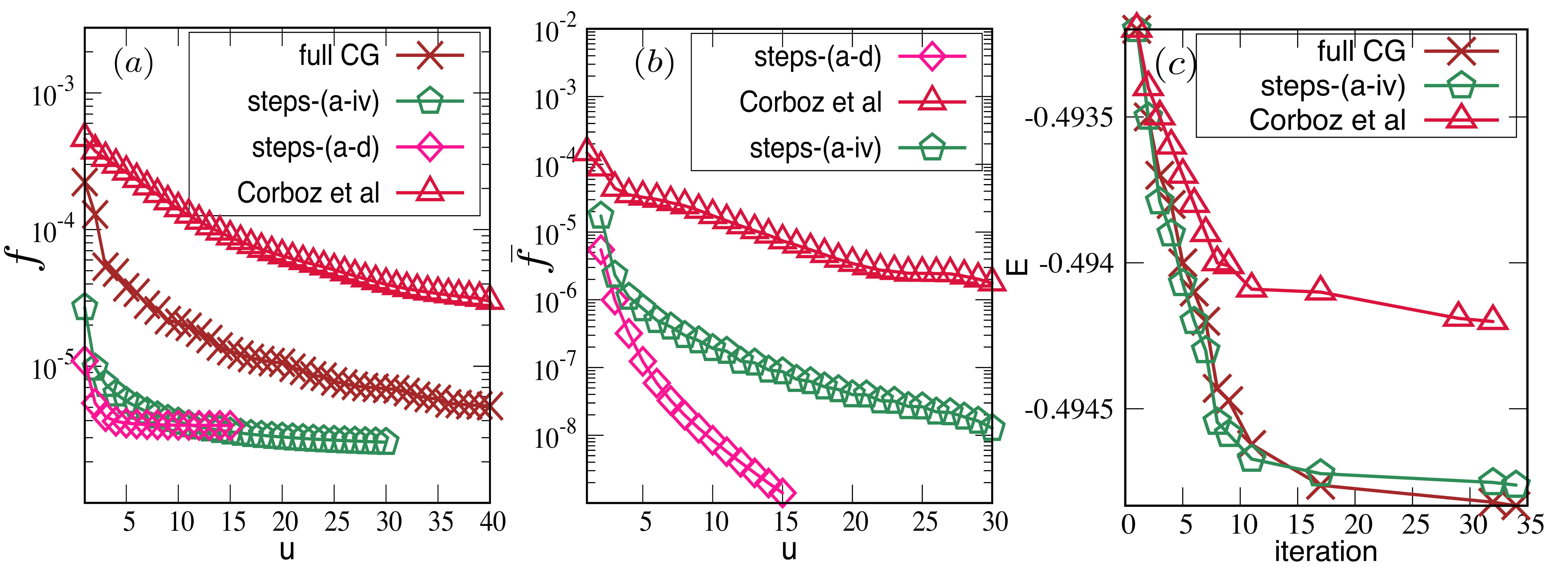}  
  \caption{(Color online) $(a, b)$ Log-linear plots of the cost function $f$ and its mean value of the relative change $\bar{f}$ as a function of consecutive iterations $u$ for different optimization schemes with bond dimension $D=6$ at $J_2={0.5}$. The data were generated for a fixed time step $\delta=0.02$. $(c)$ The ground-state energy versus loop iterations for the optimization schemes. The data shown by red triangular are obtained by the optimization method introduced in Ref.~\onlinecite{Corboz:2013}. } 
 \label{fig:Guage-opt}
\end{center}
\end{figure}

The optimization procedure is completed by iteratively repeating steps-($a$-$d$) and -($i$-$iv$) until the cost function does not change up to the desired threshold. In Fig.~\ref{fig:Guage-opt}-(a, b) we have plotted the typical behavior of the cost function $f$ and its mean value of the relative change \cite{Lubasch:2014, Phien:2015} $\bar{f}=|\frac{f_{u+1}-f_{u}}{f_{\text{init}}}|$ versus consecutive iteration number $u$ for different optimization schemes. It is seen that our scheme significantly improves convergence rate and provides better accuracy than previous schemes. \cite{Corboz:2013, Corboz:2010} In the full CG method, all tensors $\{a', b', c', d'\}$ are entirely optimized which makes its final result highly accurate. We empirically observe that CG method eventually provides better accuracy than our scheme after $ \sim 100$ iterations. In Fig.~\ref{fig:Guage-opt}-(c), we have plotted the effect of these optimization schemes on the ground-state energy. The ground-state energy is calculated at $J_{2}=0.5$ by using $U(1)$-symmetric iPEPS with bond dimension $D=6$. Similarly, it shows that our scheme improves the ground-state energy as expected.

\section{Results}\label{Sec:Results}
\subsection{Simulation remarks}
In our simulation, we run several full-update simulations initialized by random and/or ordered states (such as N\'{e}el, VBS and Stripe) to find the lowest variational ground-state energy. We first pick up the symmetric sectors with the scheme explained in Sec.~\ref{Sec:Methods} and then start the optimization procedure by performing the iterative scheme. At the end, a few steps of the full CG method is used to improve the results even more. All data points reported here correspond to the lowest variational ground-state energy that we have been able to obtain. The largest bond dimensions that we could afford are $(D,\chi)=\{(8,150), (9,100)\}$. 

We always check the behavior of the ground-state energy with respect to $\chi$ to make sure that the error due to the environment approximation is negligible. The expectation values are calculated by a modified CTM renormalization group approach.\cite{Corboz:2014, Huang:2012} We find that this approach produces much better convergence rate and more accurate results in comparison with other variants of CTM ones.\cite{Orus:2009, Orus:2012, Corboz:2010}

\subsection{Order parameters}
We need to define some order parameters to establish different ordered phases appeared in the $J_{1}-J_{2}$ SHM. Magnetically ordered phases could be addressed by magnetization parameter $m=\frac{\sum_{i} |  \langle  \textbf{S}^{z}_{i} \rangle  |}{4}$, where the index $i$ runs over the sites corresponding to the building-block tensors $\{a, b, c, d\}$. In addition, we use the local the nearest neighboring bond energy to detect the translational lattice symmetry breaking. To this end, we define the following order parameters 
\begin{align*}
\Delta T_{x}=\max(E_{x})-\min(E_{x}), \,  \Delta T_{y}=\max(E_{y})-\min(E_{y}), 
\end{align*}
where $E_{x}$ and $E_{y}$ stand for local nearest neighboring bond energy in the $x$- and $y$-directions. The order parameters $ \Delta T_{x}$ and $\Delta T_{y}$ stand for different type of lattice symmetry breaking. The lattice order parameters plus the magnetization are capable of distinguishing between the phases appeared in the $J_{1}-J_{2}$ SHM. In the N\'{e}el phase, we expect $\{ m \neq 0, \Delta T_{x}=0, \Delta T_{y}=0\}$ as the local bond energy remains the same in different directions. In the columnar VBS phase orientated in $y$-direction (analogous to one in Fig.~\ref{fig:phasediagrame}), we expect $\{ m=0, \Delta T_{x}=0, \Delta T_{y}  \neq 0\}$, while in AFM Stripe phase it becomes $\{m \neq 0, \Delta T_{x}=0, \Delta T_{y}=0\}$ (see Fig.~\ref{fig:phasediagrame}). Thus, by studying $m$ and $\Delta T_{y}$, we are able to detect N\'{e}el-to-VBS and VBS-to-Stripe quantum phase transitions.



In order to study quantum critical points, we use (connected) transverse correlation function defined by 
\begin{equation*}
C_{t}(r)=\langle \widehat{\mathcal{O}}_{(x, y)} \widehat{\mathcal{O}}_{(x+r, y)} \rangle  -  \langle \widehat{\mathcal{O}}_{(x, y)}  \rangle \langle \widehat{\mathcal{O}}_{(x+r, y)}  \rangle,
\end{equation*}
where indices $(x, y)$ show spatial coordinate and subindex $t$ stand for word `transverse'. The operators $\widehat{\mathcal{O}}_{(x, y)}$ are chosen to be $\textbf{S}_{(x, y)}$ and $\textbf{S}_{(x, y)}\cdot\textbf{S}_{(x, y+1)}$, respectively, for the spin-spin ($C_{t}^{s}$) and dimer-dimer ($C_{t}^{d}$) correlation functions. The correlation function could determine universality class of a critical phase, revealed in the power-law behavior. It algebraically falls off at critical point as 
\begin{align*}
C_{t}^{s} \sim r^{-(1+\eta_{s})}, \\
C_{t}^{d} \sim r^{-(1+\eta_{d})},
\end{align*}
where $\{\eta_{s}, \eta_{d}\}$ are anomalous spin and dimer exponents, respectively. 
A finite bond dimension $D$ (usually) induces exponential decay ($C_{t}\sim e^{\frac{-r}{\xi}}$) for large distances $r \gg 1$. Thus, we find the correlation length $\xi$ by obtaining the slopes of the following function
\begin{equation*}
\log(C_{t}(r))=(\frac{-1}{\xi})r+const \quad r \gg 1.
\end{equation*}
and obtain the critical behavior through scaling to large bond dimension limit. Note that in this method, there is one associated correlation length for each correlation function. We could also define characteristic correlation length $\xi$ by using the eigenvalues of the transfer matrix as shown in Fig.~\ref{fig:OptEXPCT}-(c). It is given by $\xi=\frac{-1}{\log(|\frac{\lambda_{2}}{\lambda_{1}}|)}$ where $\lambda_{1}$ and $\lambda_{2}$ ($|\lambda_{1}|>|\lambda_{2}|$) are respectively the largest eigenvalues of the transfer matrix. 

In some cases, when the system reveals different correlation lengths in $x$- and $y$-directions, we also need to define longitudinal correlation functions given by  
\begin{align*}
C_{l}(r)=\langle \widehat{\mathcal{O}}_{(x, y)} \widehat{\mathcal{O}}_{(x, y+r)} \rangle  -  \langle \widehat{\mathcal{O}}_{(x, y)}  \rangle \langle \widehat{\mathcal{O}}_{(x, y+r)}  \rangle.
\end{align*}
where subindex $l$ stand for word `longitudinal'.  
\subsection{N\'{e}el phase}
\label{Sec:Neelphase}

\begin{figure}
  \begin{centering}
\includegraphics[width=1.0 \linewidth]{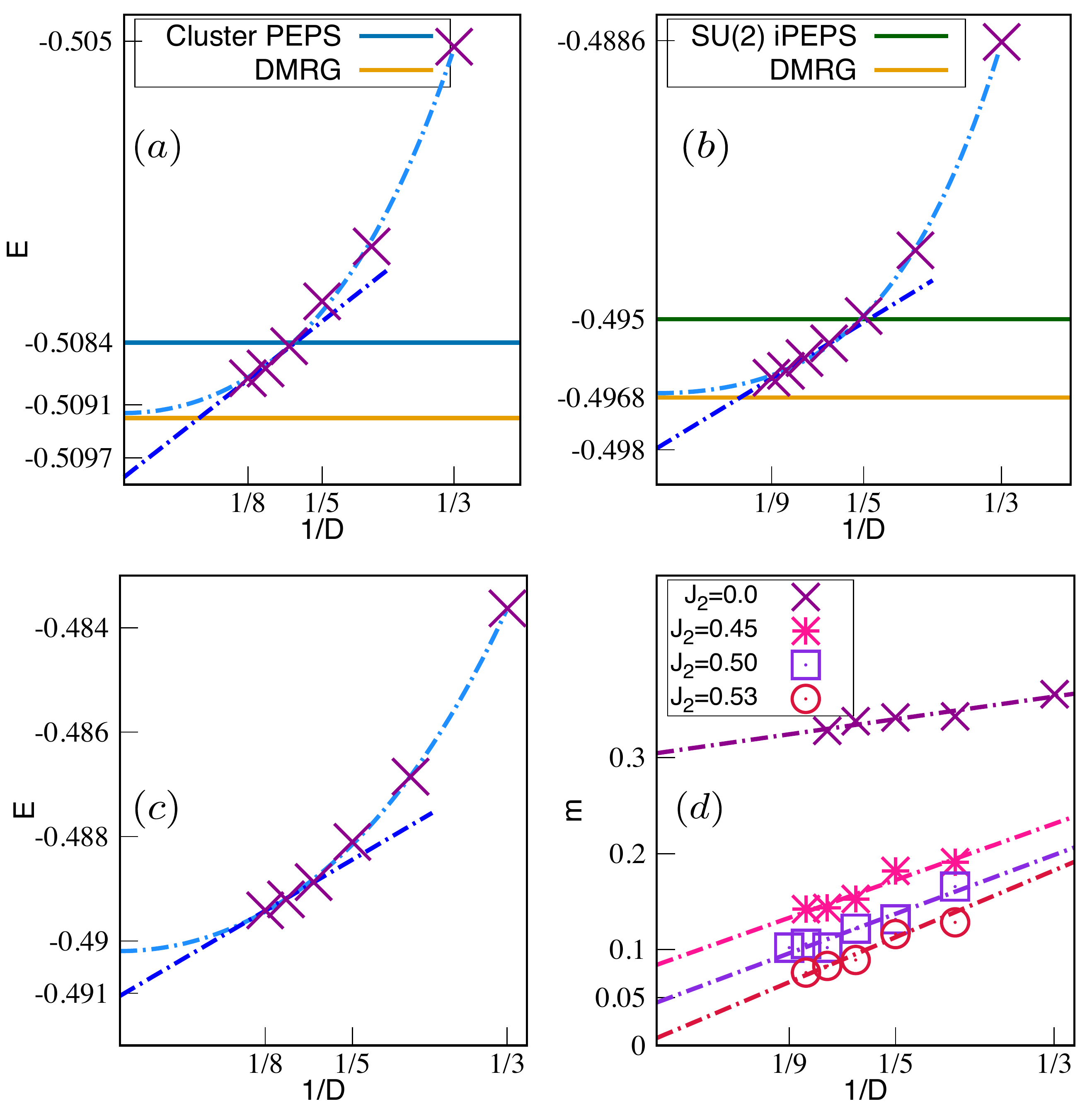}  
  \caption{(Color online) $(a, b)$ A comparison between the $U(1)$-symmetric iPEPS variational energy with that of the finite-size PEPS,\cite{Wang:2016} DMRG \cite{Gong:2014} and $SU(2)$-symmetric iPEPS \cite{Poilblanc:2017} at $J_{2}=\{0.45, 0.5 \}$, respectively. The blue (purple) dashed line represent polynomial (linear) fit up to the fourth order. $(c)$ The $U(1)$-symmetric iPEPS variational energy at the deconfined quantum-critical point $J^{c_1}_{2}=0.53$. $(d)$  The N\'{e}el order parameter $m$ as a function of $1/D$. A linear fit in large-D limit reveals $m$ vanishes at point $J_{2}=0.53$.}
  \label{fig:NeelEnergy}
\end{centering}
\end{figure}

In Fig.~\ref{fig:NeelEnergy}-(a), we have compared the ground-state energy obtained by $U(1)$-symmetric iPEPS with the extrapolated value ($ N \rightarrow \infty $) of the finite-size PEPS \cite{Wang:2016} at $J_{2}=0.45$. Our ground-state energies for bond dimensions $D\geq6$ are lower than that of the finite-size PEPS with a larger bond dimension $D=9$. The PEPS results are obtained by the cluster-update scheme \cite{Lubasch:2014} which is considered as an intermediate optimization approach between simple and full update---it is computationally cheaper than full update. Our best variational energy $E_{\text{iPEPS}}^{D=8}=-0.5088$ sets an upper bound to the true ground-state energy at $J_{2}=0.45$. 

We have also compared the result of the $SU(2)$-symmetric iPEPS ansatz \cite{Poilblanc:2017} applied recently to the $J_{1}-J_{2}$ SHM at $J_{2}=0.5$. As seen in Fig.~\ref{fig:NeelEnergy}-(b), $U(1)$ iPEPS with bond dimensions $(D,\chi)=(5, 40)$ provides the same ground-state energy as $SU(2)$ iPEPS with bond dimensions $(D,\chi)=(7, \chi \rightarrow \infty)$. The reason might be due to the effect of finite bond dimension; as the $SU(2)$ iPEPS provides an efficient representation for only symmetric phases. We argue that the system at $J_{2}=0.5$ is still magnetically ordered, thus, the $SU(2)$ iPEPS probably picks up a superposition of the states requiring larger bond dimensions. Our best variational energy at the highly frustrated point $J_{2}=0.5$ is $E_{\text{iPEPS}}^{D=9}=-0.4964$, which is quite close to the DMRG extrapolated value, $E_{ \text{DMRG}}=-0.4968$. \cite{Gong:2014}

We study the N\'{e}el order parameter $m$ as a function of the bond dimension $D$ to find the critical point, where the N\'{e}el phase disappears. At point $J_{2}=0$, we find that a linear extrapolation with the large bond dimensions ($D \geq 4$) provides a proper estimation of $m$. The relative error of our estmation with that of the Monte-Carlo result~\cite{Sandvik:1997} is of order $\Delta m= \frac{m_{D \rightarrow \infty}-m_{\text{MC}}}{m_{\text{MC}}}<3\times10^{-2}$. In Fig.~\ref{fig:NeelEnergy}-(d), we have plotted the N\'{e}el order parameter $m$ versus $1/D$ for different values of $J_{2}$. A linear extrapolation (dashed lines) for the larger bond dimensions ($D\geq 4$) reveals that $m$ remains finite in the range of $0 \leq J_{2} \leq 0.53$. In this interval, the order parameters $\Delta T_{x}<0.03$ and $\Delta T_{y}<0.03$ are both small consistent with that the N\'{e}el phase persists up to point $J_{2}=0.53$. At this point, $m$ is almost zero ($<10^{-4}$), thus, we conclude the quantum critical point occurs at $J^{c_1}_{2}=0.530(5)$. The ground-state energy at this point has been shown in Fig.~\ref{fig:NeelEnergy}-(c): the best upper bond on the ground-state energy and the extrapolated value are $E_{\text{iPEPS}}^{D=8}=-0.4894$ and $E_{\text{iPEPS}}^{D \rightarrow \infty}=-0.4902$ (from polynomial fit), respectively.

\subsection{Columnar VBS phase}
\label{Sec:ColumnarVBSphase}
\begin{figure}[b]
\begin{centering}
\includegraphics[width=1.0 \linewidth]{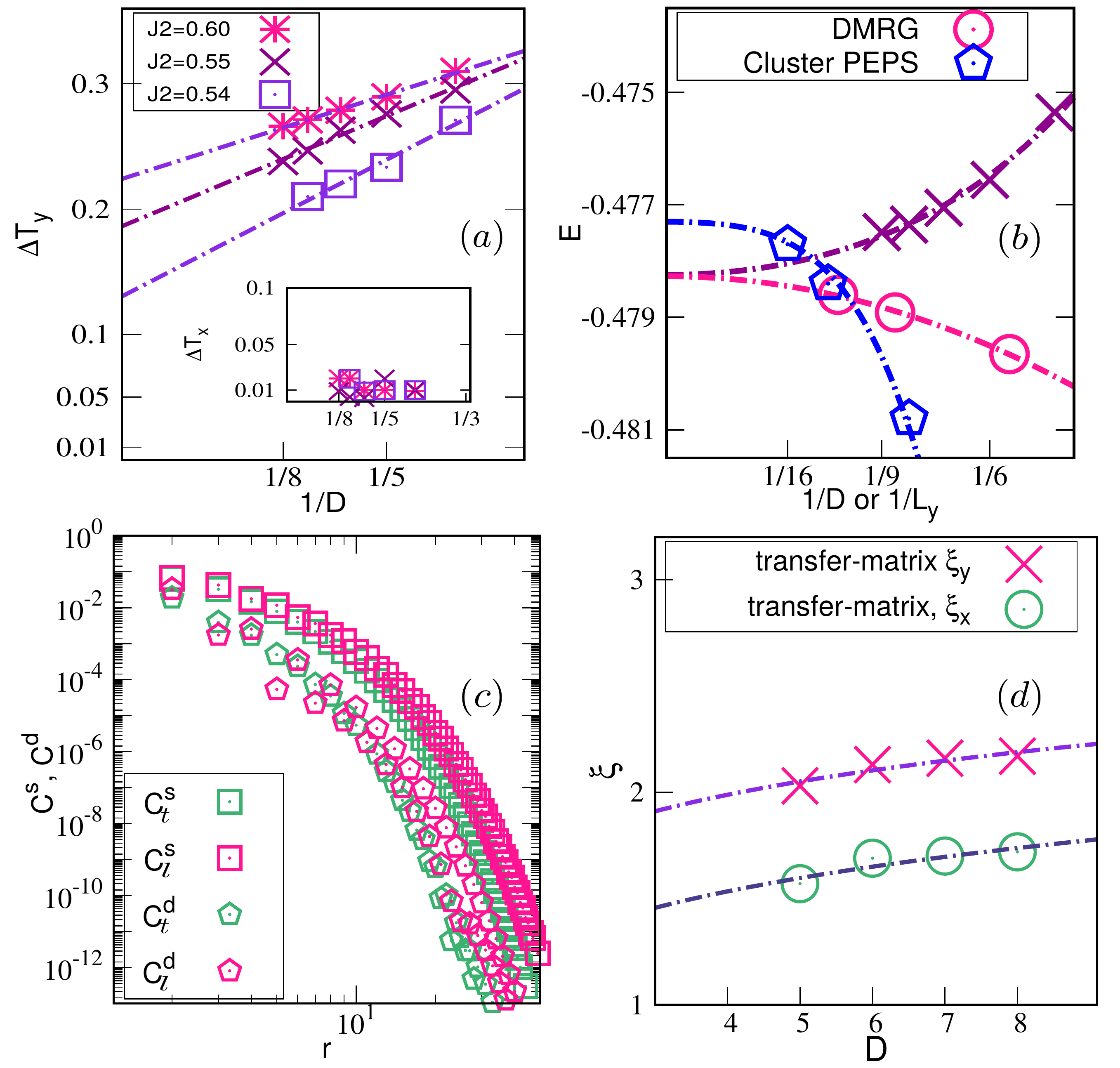} 
  \caption{(Color online) $(a)$ Order parameters $\Delta T_{y}$ and $\Delta T_{x}$ (inset) as a function of bond dimension $1/D$. $(b)$ The variational ground-state energies at $J_{2}=0.6$. The DMRG and the finite-size PEPS data are respectively obtained on the tilted cylinder with width $L_{y}$ and $L_y\times L_y$ torus. $(c)$ Log-log plot of the spin-spin and dimer-dimer correlation functions versus distance $r$ in the $x$- and $y$-directions. The date has been plotted for bond dimension $D=8$, $(d)$ The correlation length versus bond dimension $D$. The dashed lines are power-law fits $\xi \sim D^{\alpha}$.}
\label{fig:CVBS} 
\end{centering}
\end{figure}

We study the order parameters $\Delta T_{x}$, $\Delta T_{y}$ and correlation functions for $J_{2}>0.53$ to find the true nature of the non-magnetic phase. We plot $\Delta T_{x}$ and $\Delta T_{y}$ for points $J_{2}=\{0.54, 0.55, 0.6\}$ as depicted in Fig.~\ref{fig:CVBS}-(a). It suggests a columnar VBS order for no-magnetic phase: in the large-D limit, $\Delta T_{y}$ remains finite, while $\Delta T_{x} $ is one order of magnitude smaller than $\Delta T_y$. By increasing $J_{2}$, the order parameter $\Delta T_{y}$ monotonically increases and reaches its maximum value around $J_{2} \approx 0.61$---we will show later that a first-order phase transition takes place at this point. To check the validity of the result, we compare the ground-state energy with previous studies at $J_{2}=0.6$, as depicted in Fig.~\ref{fig:CVBS}-(b). At this point, DMRG\cite{Gong:2014} and finite-size PEPS\cite{Wang:2016} study respectively predicted a plaquette VBS order and a critical behavior (algebraic fall-off of the correlation function up to $N=24 \times 24$). We expect that this critical behavior, in the PEPS calculation, eventually disappears in the thermodynamic limit. We notice that the essential difference between the iPEPS and PEPS anstaz lies in the finite-size boundary effects, as both are using the same underlying tensor-network wave function. Since our variational energy is quite compatible with that of finite-size PEPS and our result predicts a VBS order, we might conclude that as the system size increases, algebraic fall-off of the correlation function get eventually dominated by an exponential behavior.

In order to gain more insight, we investigate the correlation functions at the point $J_{2}=0.6$. In Fig.~\ref{fig:CVBS}-(c), we have plotted the the transverse and longitudinal correlation functions. The longitudinal and transverse correlation functions show different correlation lengths as expected from the nature of the columnar VBS ordered state. We observe that they exponentially fall off, as confirmed by the behavior of correlation lengths, shown in Fig.~\ref{fig:CVBS}-(d). The characteristic correlation lengths $\xi_{x, y}$ increase slowly with bond dimension $D$ and seem to saturate in the large-$D$ limit. A power-law fit $\xi \sim D^{\alpha}$ to the largest bond dimensions $D=\{6, 7, 8\}$ reveals $\alpha<0.07$.

\begin{figure}
\begin{centering}
\includegraphics[width=1.0 \linewidth]{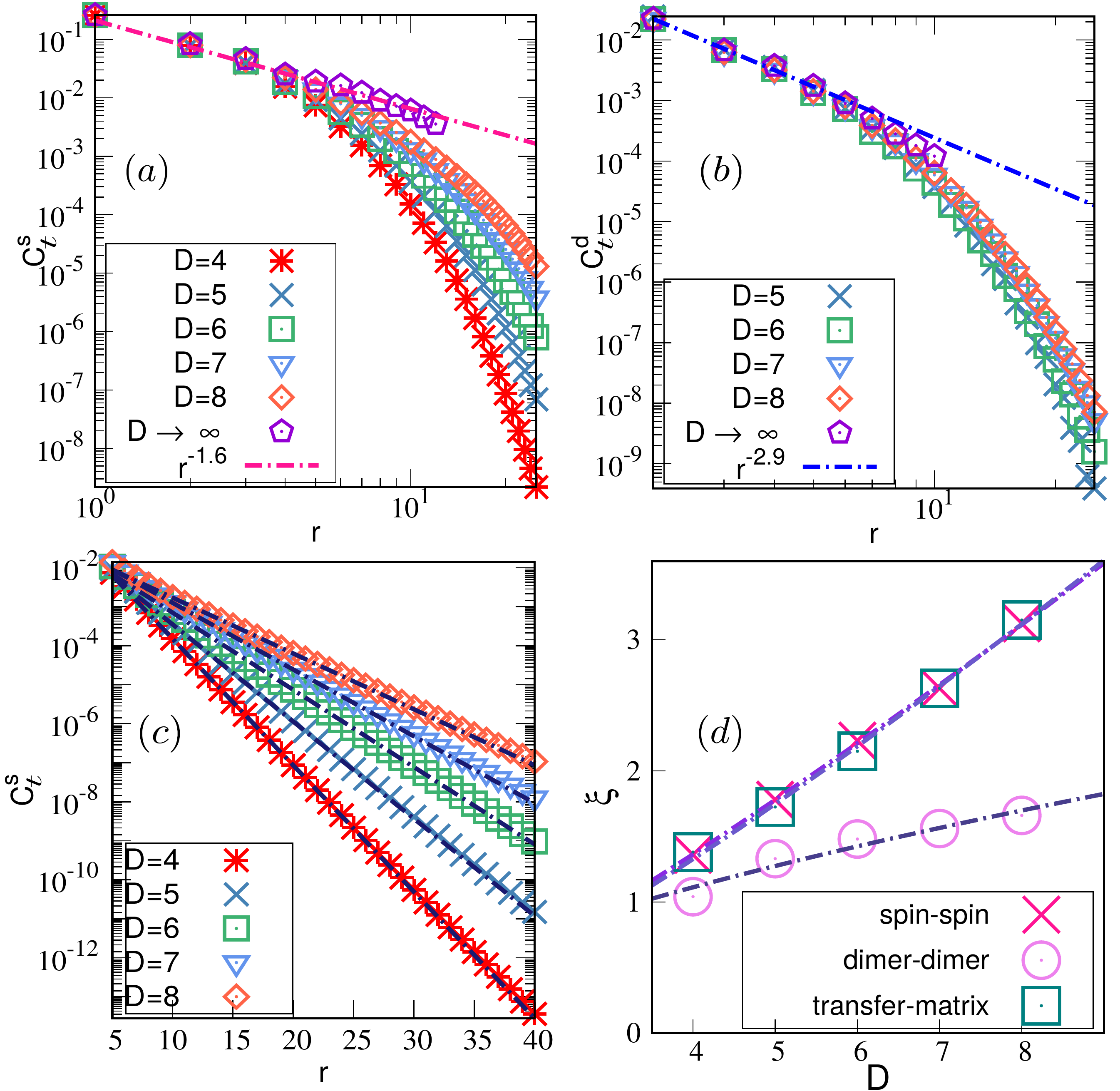} 
  \caption{(Color online) Behavior of the correlation functions at deconfined quantum-critical point . $(a, b)$ Log-log plot of the spin-spin and dimer-dimer correlation functions versus distance $r$. $(c)$ Log-linear plots of the spin-spin correlation function versus distance $r$, where slopes show inverse of the correlation length $\xi^{-1}_{s}$. $(d)$ The correlation lengths as a function of bond dimension $D$; dashed lines are power-law fits $\xi \sim D^{\alpha}$.}
\label{fig:Corr} 
\end{centering}
\end{figure}

\subsection{Deconfined quantum criticality}
\label{Sec:Deconfined}
In this section, we investigate critical properties of the deconfined quantum-critical point by studying the correlation functions and the associated correlation lengths. We study correlation functions at the critical point $J_{2} = 0.53$ and compare the results with the previous studies. In Fig.~\ref{fig:Corr}-(a, b), we have plotted correlation functions $C_{t}^{s}$ and $C_{t}^{d}$ as a function of distance $r$. The data for each bond dimension $D$ are obtained by the largest environment bond dimension $\chi$, although in contrast to Ref.~\onlinecite{Poilblanc:2017} we do not observe any strong dependency on $\chi$. In order to understand the true behavior of the correlation functions, we need to study the associated correlation lengths as a function of $D$. In Fig.~\ref{fig:Corr}-(c), we show the log-linear plot of the spin-spin correlation function versus large distance $r \gg 1$. The slopes reveal the inverse of the spin correlation length $\xi^{-1}_{s}$. $\xi_{s}$ increases significantly by increasing the bond-dimension $D$ as expected in a critical regime. They follow an empirical power-law relation $\xi \sim D^{\alpha}$ as shown in Fig.~\ref{fig:Corr}-(d). The spin correlation length $\xi_{s}$ and characteristic correlation length $\xi$ (extracted from the transfer matrix) diverges similarly as $\xi_{s}, \xi \sim D^{1.2}$. Instead, dimer correlation length is governed by different scaling exponent as $\xi_{d} \sim D^{0.5}$.\cite{Corr-note} This divergent behaviors suggest that $J_{2} \approx 0.53$ is a critical point which is consistant with vanishing the order parameters.\cite{iPEPS-note}

The divergent behavior of correlation length $\xi(D)$ implies an algebraic fall-off of the correlation function in the range of $1<r<\xi(D)$. An accurate estimation of the critical anomalous exponents requires $\xi(D)$ to be large enough. Particularly, in our case, $\xi(D)$ is still small even for the largest bond dimension. Thus, we need to rely on the extrapolated data in the $D \rightarrow \infty$ limit, which correspondingly represent a large correlation length $\xi(D \rightarrow \infty)>>1$. We use a linear extrapolation $D>4$ to obtain the extrapolated data of correlation function up to $r \sim 12$, where error-bars are still small (see Appendix.~\ref{App:Extrapolated}). As shown in in Fig.~\ref{fig:Corr}-(a, b), we have fitted the data (in $D\rightarrow \infty$) to a power-law function $r^{-(1+\eta)}$ to estimate the exponents. The critical exponents for spin-spin and dimer-dime correlations are, respectively, $\eta_{s} \sim 0.6$ and $\eta_{d} \sim 1.9$, which are in agreement with Ref.~\onlinecite{Poilblanc:2017}. Our results show that dimer-dimer correlation falls off more rapidly than predicted by $\mathcal{J}$-$\mathcal{Q}$ models ($0.26<\eta_{\text{$\mathcal{J}$-$\mathcal{Q}$}}<0.6$). \cite{Sandvik:2012, Sandvik:2009} This may indicate different universality classes of the deconfined criticality for different models.

Therefore, our results predict a continuous N\'{e}el-to-VBS transition, which is forbidden in Landau-Ginzburg theory due to the different types of broken symmetry---unless it would be of the first-order type. So, we conclude that this quantum phase transition fits well in the paradigm of `deconfined quantum criticality'.\cite{Senthil:2004} However, the field-theory description of this deconfined quantum critical point might be different from that of the $\mathcal{J}$-$\mathcal{Q}$ models, as seen by different scaling behavior of correlation functions.


\subsection{First-order quantum phase transition}
\label{Sec:FirstOrder}
\begin{figure}
\begin{centering}
\includegraphics[width=1.0 \linewidth]{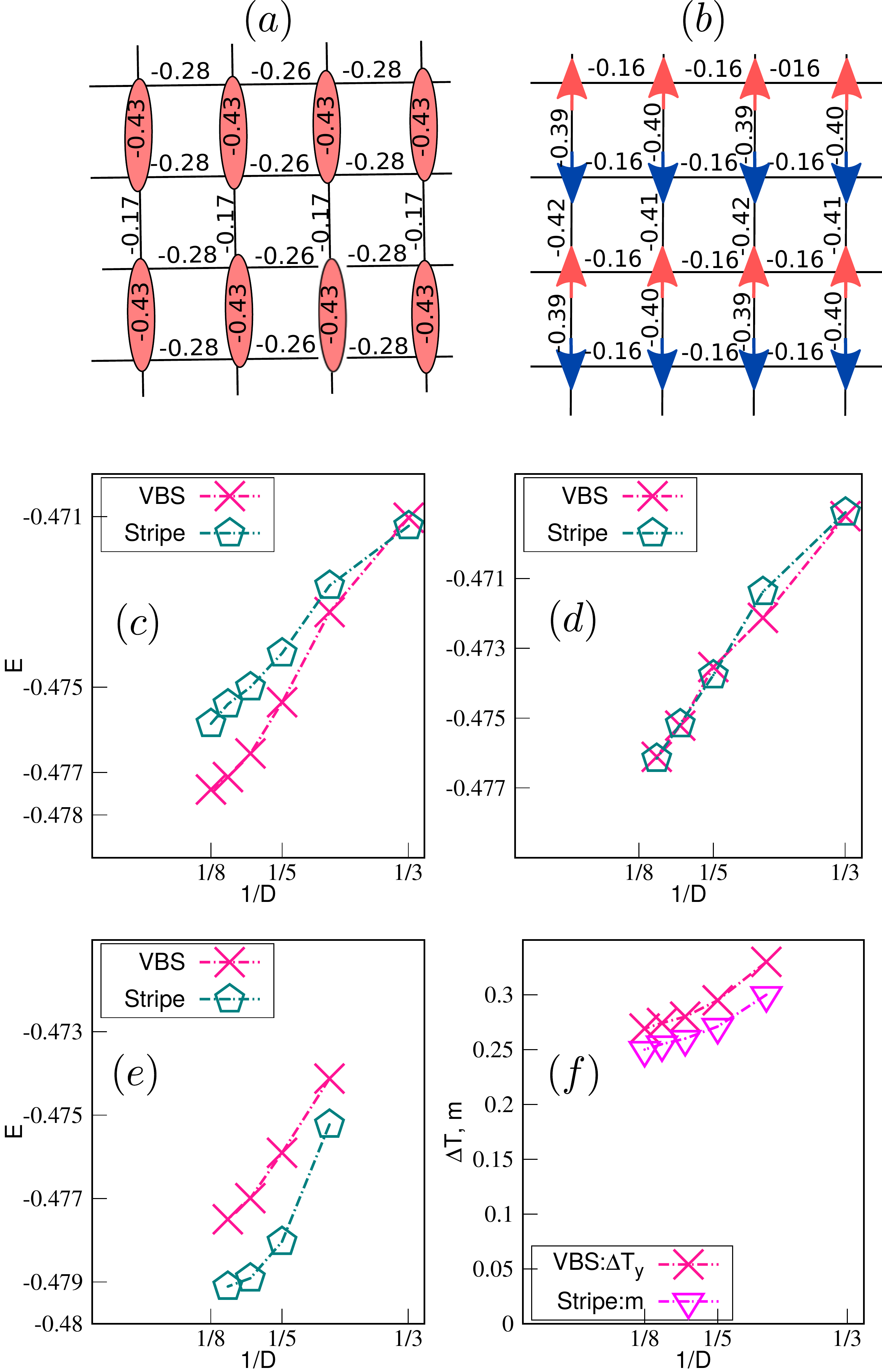} 
  \caption{(Color online) $(a, b)$ Schematic pattern of $J_{1}$ bond-energy at $J_{2}=0.6$ and $J_{2}=0.62$ for bond dimension $D=8$. $(c, d, e)$ The $U(1)$-symmetric iPEPS ground-state energy initialized by N\'{e}el and Stripe states at $J_{2}=\{0.60, 0.61, 0.62\}$, respectively. $(d)$ The order parameters $\Delta T_{y}$ and $m$ at the point $J_{2}=0.61$. It shows both N\'{e}el and AFM Stripe states exists at this point.}
\label{fig:bond-energy}  
\end{centering}
\end{figure}

We expect a quantum phase transition to occur between columnar VBS and AFM Stripe phases as $J_{2}$ increases. To locate the quantum phase transition point, we sketch diagrammatically local nearest neighboring ($J_{1}$) bond energy at $J_{2}=\{0.60, 0.62\}$, see Fig.~\ref{fig:bond-energy}-(a, b). The pattern of bond energy shows the lattice symmetry breaking in the $y$-direction ($\Delta T_{y} \sim 0.2$ and $\Delta T_{x} \sim 0.01$) disappears at $J_{2}=0.62$, where the AFM Stripe phase emerges. Since order parameter $\Delta T_{y}$ has been monotonically increased from the point $J_{2}=0.54$, we expect the quantum phase transition to be the first-order type rather than continuous one.

We use hysteresis analysis as explained in Ref.~\onlinecite{Corboz:2013} to find whether the quantum phase transition is the first-order type: $(i)$ we initialize the iPEPS ansatz by competitive ordered states in the vicinity of the critical point, $(ii)$ find where the energies become equal for different values of bond dimension $D$ and $(iii)$ check whether the order parameters remain non-zero. As shown in Fig.~\ref{fig:bond-energy}-(c, d, e), we have compared the energies internalized by columnar VBS  and AFM Stripe states at $J_{2}=\{0.60, 0.61, 0.62\}$. We observe that energies of states with different initializations at the point $J_{2}=0.61$ become almost equal, but for $J_{2}=\{0.60, 0.62\}$, the columnar VBS and the AFM Stripe respectively provide lower energy. Thus, we conclude that they cross around $J_{2}=0.61$. At this point, the order parameters for both states remain finite, as shown in in Fig.~\ref{fig:bond-energy}-(f). For columnar VBS  and AFM Stripe, we respectively obtain in the large-$D$ limit $(\Delta T_{y}, \Delta T_{x}) \approx (0.24, 0.01)$ and $ m \approx 0.21$. Therefore, the transition occurring $J^{c_2}_{2}=0.610(3)$ is of the first order.

\section{DISCUSSION AND CONCLUSION}
\label{Sec:CONCLUSION}
In this paper, we have addressed two main obstacles regarding the iPEPS ansatz: $(i)$ how to improve iPEPS update schemes in the presence of second-neighbor interactions \cite{Corboz:2010} and $(ii)$ how to automatically select relevant symmetry sectors (in the case of continuous symmetry) without losing accuracy.\cite{Bauer:2011} We considered the first issue by introducing an `improved' update scheme based on positive approximant and reduced-tensor application. The update scheme significantly accelerates the convergence rate and also improves the accuracy/stability in comparison with previous schemes. \cite{Corboz:2010:April, Corboz:2013} For the second issue, a simple strategy is introduced to pick up relevant symmetry sectors so that the accuracy remains the same as non-symmetric cases. We also showed that taking a non-homogeneous structure for all virtual bonds is crucial in the case of the $U(1)$-symmetric iPEPS ansatz---which does not seem to be the case for finite symmetry groups.

We utilize our $U(1)$-symmetric iPEPS ansatz to investigate the ground-state phase diagram of the $J_{1}-J_{2}$ SHM on the square lattice. A N\'{e}el phase is found for $J_2 < 0.53$ by observing a non-zero value of the magnetically order parameter in the large-D limit. In the range $0.53<J_{2}<0.61$, by studying the lattice symmetry breaking order parameters, we find that a columnar VBS phase is established. The point $J^{c_1}_2=0.53$ represents a deconfined N\'{e}el-VBS quantum critical point, as confirmed by vanishing the order parameters and divergent behavior of the characteristic correlation length and spin correlation length, i.e., $\xi \sim D^{1.2}$. This result is consistent with that of DMRG studies: accurate $SU(2)$-symmetric DMRG \cite{Gong:2014} estimates the transition point $\approx > 0.50$, while a very recent $U(1)$-symmetric DMRG study \cite{Wang:2017} based on level spectroscopy has predicted the transition to be $\approx 0.52$, although a small window of possible gapless spin-liquid is suggested in this work. Our findings improve the result of finite-size PEPS study \cite{Wang:2016} which obtained a critical point around $\approx 0.57$. The main reason for such difference may come from the lack of the finite-$D$ extrapolation in Ref.~\onlinecite{Wang:2016}.  

We have studied dimer-dimer and spin-spin correlation functions to compare the associated critical exponents with that of the $\mathcal{J}$-$\mathcal{Q}$ model, i.e. $\eta_{\text{$\mathcal{J}$-$\mathcal{Q}$}} \sim 0.26$. Our estimated dimer and spin anomalous exponents, $\eta_{s} \sim 0.6$ and $\eta_{d} \sim 1.9$, show deviation from that value. That observation is also manifested in the divergent behavior of the correlation lengths: the spin and dimer correlation lengths diverge as $\xi_{s} \sim D^{1.2}$ and $\xi_{d} \sim D^{0.5}$, respectively. A very recent $SU(2)$-iPEPS study \cite{Poilblanc:2017} has suggested that spin correlation length diverges linearly with environment bond dimension $\chi$, $\xi_{s} \sim \chi$ (although, in contrast, we do not observe any strong dependency on $\chi$ in our calculations). 

The pattern of local nearest neighboring bond energy reveals that the nature of the VBS order is of the columnar type. The associated VBS order parameter increases monotonically up to the point $J_{2} \approx 0.61$, where a first-order phase transition occurs. In comparison with the plaquette VBS order predicted by DMRG simulations, both phases seem to be quite competitive. We have estimated transition point at $J_{2} = 0.610(3)$ based on hysteresis analysis. At this point both associated order parameters of the columnar VBS and the AFM Stripe are non-zero.

Our study clearly shows that the iPEPS ansatz finds a non-zero N\'{e}el order parameter in the range of $0.45<~ J_2<~0.5$, where DMRG studies predict a possible gapless phase. It is an interesting direction to improve both methods further to obtain more accurate estimation of the relevant order parameters and reach a rigorous conclusion for that phase. A natural next step is to apply the method, determining relevant symmetric sectors, to the $SU(2)$-symmetric iPEPS ansatz, which might improve the accuracy similar to the $U(1)$ case. In addition, for models with long-range interactions, such as $J_{1}-J_{2}-J_{3}$ Heisenberg models, an efficient generalization of the update scheme is needed. Furthermore, using the $U(1)$-symmetric iPEPS ansatz for larger-spin systems (defined on different geometries) to characterize different quantum phases is another direction of further studies. 

\acknowledgments
We thank Shou-Shu Gong for stimulating discussions. We also acknowledge Mac Lee for reading the manuscript. This research is supported by National Science Foundation Grants PREM DMR-1205734 and DMR-1408560. We use \emph{Uni10 library}, \cite{Kao:2015} an open-source library, to build and perform the iPEPS algorithms introduced in this paper.

\bibliography{references} 

\begin{thebibliography}{64}
\bibitem [{Cor()}]{Corr-note}%
  \BibitemOpen
  \href@noop {} {}\bibinfo {note} {Our results support a divergent behavior for
  the dimer correlation length at the critical point $J_{2} \sim 0.53$.
  However, in order to obtain a more rigorous conclusion, to show that it does
  not tend to a finite value in the $D \rightarrow \infty$ limit, the larger
  bond dimension is required, which is beyond our current
  capability.}\BibitemShut {Stop}%
\bibitem [{iPE()}]{iPEPS-note}%
  \BibitemOpen
  \href@noop {} {}\bibinfo {note} {The typical behavior of correlation
  function, critical exponents and correlation length are almost the same in
  the range $J_{2} \sim 0.5-0.53$. So, we mainly rely on vanishing N\'{e}el
  order parameter to estimate the critical point.}\BibitemShut {Stop}%
\end{thebibliography}
\begin{figure}
\begin{center}
\includegraphics[width=1.0 \linewidth]{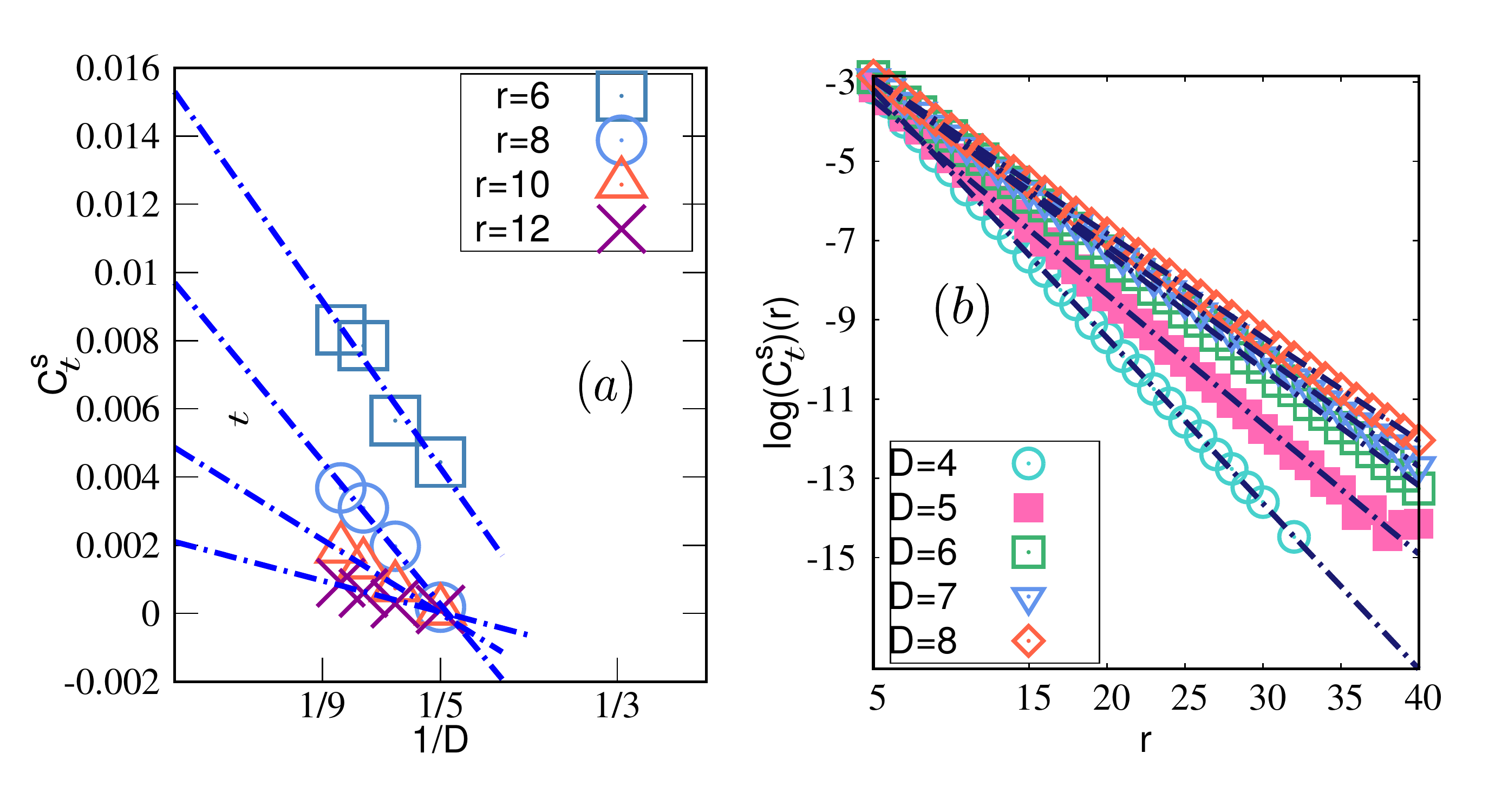}  
  \caption{(Color online) The extrapolated data points for the correlation function. $(a)$  The spin-spin correlation function versus bond dimension $1/D$ for different value of $r$. A linear fit is used to estimate the extrapolate data in the $D\rightarrow \infty$. $(b)$ Log-linear plot of the dimer-dimer correlation function versus distance $r$, where slopes show inverse of the correlation length $\xi^{-1}_{d}$. } 
\label{fig:Fitting}
\end{center}
\end{figure}

 \appendix
\section{Extrapolated data for the correlation functions}
\label{App:Extrapolated} 
 In this section, we provide further data points of the correlation functions and discuss the extrapolation procedure used in the estimation of the critical exponents. In order to estimate, e.g., the spin critical exponent, we first obtain the spin-spin correlation function $C_{t}^{s}(r)$ for the large bond dimensions $D \sim 5-8$. Then, we use a linear fit (in $1/D$) to extrapolate $C_{t}^{s}(r)$ in the $D\rightarrow \infty$ limit; As depicted in Fig.~\ref{fig:Fitting}-(a), we have plotted $C_{t}^{s}(r)$ as a function of $1/D$ and have shown the linear fits for different values of distance $r\sim 6-12$. A linear fit seems to provide reliable estimation of the extrapolated data points. We finally use the these data points to estimate the exponents as shown in Fig.~\ref{fig:Corr}-(a). 
 
We also report in Fig.~\ref{fig:Fitting}-(b), the log-linear plot of the dimer-dimer correlation function versus large distance $r \gg 1$. The slopes reveal the inverse of the spin correlation length $\xi^{-1}_{d}$, which seems to weakly depend on $D$. We use the slopes in Fig.~\ref{fig:Corr}-(d) to study D-dependence behavior of $\xi_{d}$.

\end{document}